\documentclass[useAMS,usenatbib]{mn2e}
\usepackage[dvips]{graphicx}
\title[Hot nights on extrasolar planets]{Hot nights on extrasolar planets:\\mid-infrared phase variations of hot Jupiters}
\author[N.B. Cowan, E. Agol and D. Charbonneau]{N.~B.~Cowan$^1$\thanks{E-mail: cowan@astro.washington.edu}, E.~Agol$^1$, D.~Charbonneau$^2$\thanks{Alfred P. Sloan Research Fellow}\\
$^1$Department of Astronomy, University of Washington, Box 351580, Seattle, WA 98195\\
$^2$Harvard-Smithsonian Center for Astrophysics, 60 Garden Street, Cambridge, MA 02138}

\def\LaTeX{L\kern-.36em\raise.3ex\hbox{a}\kern-.15em
    T\kern-.1667em\lower.7ex\hbox{E}\kern-.125emX}

\label{firstpage}
\begin{document}

\maketitle

\begin{abstract}
We present results from Spitzer Space Telescope observations of the mid-infrared phase variations of three short-period extrasolar planetary systems: HD~209458, HD~179949 and 51~Peg. 
We gathered Infrared Array Camera (IRAC) images in multiple wavebands (3.6~$\mu$m or 4.5~$\mu$m, and 8~$\mu$m) at eight phases of each planet's orbit. We find the uncertainty in relative photometry from one epoch to the next to be significantly larger than the photon counting error at 3.6~$\mu$m and 4.5~$\mu$m. We are able to place $2\sigma$ upper limits of only $\sim 2$\% on the phase variations at these wavelengths. At 8~$\mu$m the epoch-to-epoch systematic uncertainty is comparable to the photon counting noise and
we detect a phase function for HD~179949 which is in phase with the planet's orbit and with a relative peak-to-trough amplitude of $0.00141(33)$. Assuming that HD~179949b has a radius $R_{J} < R_{p} < 1.2R_{J}$, it must  
recirculate less than $21$\% of incident stellar energy to its night side at the $1\sigma$ level (less than $26$\% at the $2\sigma$ level, where $50$\% signifies full recirculation). If the planet has a small Bond albedo, it must have a mass less than $2.4 M_{J}$ ($1\sigma$).
We do not detect phase variations for the other two systems but we do place the following $2\sigma$ upper limits: $0.0007$ for 51~Peg, and $0.0015$ for HD~209458. Due to its edge-on configuration, the upper limit for HD~209458 translates, with appropriate assumptions about Bond albedo, into a lower limit on the recirculation occuring in the planet's atmosphere.  HD~209458b must recirculate at least $32$\% of incident stellar energy to its night side, at the $1\sigma$ level (at least $16$\% at the $2\sigma$ level), which is consistent with other constraints on recirculation from the depth of secondary eclipse depth at 8~$\mu$m and the low optical albedo. 
These data indicate that different Hot Jupiter planets may experience different recirculation efficiencies.
\end{abstract}  


\section{Background}
\subsection{Introduction}
Since the discovery of the first Hot Jupiter system \citep{1995Natur.378..355M} and the subsequent realization that such systems are ubiquitous \citep{2006ApJ...646..505B}, astronomers have struggled to observe these planets and to compare these observations to theoretical predictions about their surface temperature variations, composition, albedo, and internal structure. The observation of extrasolar planetary transits, starting with HD~209458b \citep{2000ApJ...529L..45C,2000ApJ...529L..41H}, have allowed for the determination of the mass and size of these planets, but it was only with the observation of secondary eclipses, starting with HD~209458b \citep{2005Natur.434..740D} and TrES-1 \citep{2005ApJ...626..523C}, that it has been possible to measure the flux from these planets \citep[for a review of transiting exoplanet science see][]{2007PPV...701}. The depth of the secondary eclipse, however, only characterizes the day-side flux from the planet.  To fully characterize the planet's longitudinal temperature profile, observations must be made at a variety of planetary phases. The first such observations were reported by \cite{2006Sci...314..623H} and \cite{2007_Knutson}.
In this paper we present similar observations of three other planetary systems.  We briefly review models of Hot Jupiter phase variations in \S1.2 and \S1.3, and the observational constraints in \S1.4. In \S2 we present our observations and in \S3 we explain how they constrain the properties of Hot Jupiters. We state our conclusions in \S4.

\subsection{Toy Model} \label{toy}
For the purposes of this paper, the phase function of a planetary system is taken to be the relative flux contribution from the planet as a function of its orbital position. Here we describe a toy model of Hot Jupiter phase functions, based on a planet in a circular orbit with two isothermal hemispheres and neglecting the planet's remnant heat of formation, an effective temperature of $T_{\rm form} \approx$ 50--75~K \citep{2006ApJ...650.1140B}. The Bond albedo, $A_{B}$, is the fraction of the incident stellar flux that is reflected by the planet. A planet's energy budget is largely determined by its Bond albedo. Hence, if we treat the host star as a blackbody, the equilibrium temperature of a planet's day and night faces is, respectively
\begin{equation}
T_{\rm day}^{4} =  (1-A_{B})(1-P_{n}) \left( \frac{R_{*}^{2}}{2a^{2}} \right)  T_{\rm eff}^{4},
\end{equation}
and
\begin{equation}
T_{\rm night}^{4} = (1-A_{B})P_{n} \left( \frac{R_{*}^{2}}{2a^{2}} \right)  T_{\rm eff}^{4},
\end{equation}
where $R_{*}$ and $T_{\rm eff}$ are the star's radius and effective temperature, $a$ is the planet's semi-major axis and $P_{n}$ quantifies the portion of the absorbed stellar energy advected to the planet's night-side ($P_{n}=0$ for no redistribution; $P_{n}=0.5$ for full redistribution) \citep{2006ApJ...650.1140B}.

The depth of secondary eclipse (for edge-on systems) is approximately the ratio of the planet's day-side flux, $F_{\rm day}$, to the stellar flux, $F_{*}$. In light of the models of \cite{2006ApJ...652..746F}, we assume that the planet's day and night-side radiate as blackbodies ($F=B_{\lambda}(T)$) and that the host star emits with a mid-IR brightness temperature $T_{\rm bright}$ \citep[we use $T_{\rm bright}=0.87T_{\rm eff}$ at 8~$\mu$m based on the models of][]{Kurucz}. The resulting planet-star flux ratio is:
\begin{equation}
\frac{F_{\rm day}}{F_{*}} =  A_{g} \left( \frac{R_{p}}{a}\right)^{2}+ \frac{B_{\lambda}(T_{\rm day})}{B_{\lambda}(T_{\rm bright})} \left( \frac{R_{p}}{R_{*}}\right)^{2},
\end{equation}
where $R_{p}$ is the planetary radius and the ratios $R_{p}/a$ and $R_{p}/R_{*}$ can be well-constrained if the planet transits its host star. The wavelength-dependent geometric albedo, $A_{g}$, 
is the ratio of the total brightness of the planet at full phase to that of a Lambertian disk with the same cross-section.
The second term (thermal emission from the planet) can be neglected at optical wavelengths.

In the infrared, we may neglect the contribution from reflected starlight.
The peak-to-trough amplitude of the phase function for a transiting
systems then becomes:
\begin{equation}
\frac{F_{\rm day}-F_{\rm night}}{F_{*}} = \frac{B_{\lambda}(T_{\rm day})-B_{\lambda}(T_{\rm night})}{B_{\lambda}(T_{\rm bright})} \left( \frac{R_{p}}{R_{*}}\right)^{2},
\end{equation}
where $F_{\rm night}$ is the flux from the planet's night-side and where we have neglected the planet's flux contribution ($<1$\%) in the denominator. 
Note that the absolute flux from the planet is a minimum during the secondary eclipse, irrespective of phase variations, since the planet is entirely hidden from view at that time. The normalized phase variations, $\Delta F(\phi)/\langle F\rangle$, for a system with inclination $i$ is:
\begin{equation} \label{phase}
\frac{\Delta F(\phi)}{\langle F\rangle} = -\sin{i} \cos{\phi} ~\frac{F_{\rm day}-F_{\rm night}}{2F_{*}},
\end{equation}
where the $\phi$ is the planet's orbital phase ($\phi = 0$ at inferior conjunction; $\phi = \pi$ at superior conjunction) and $\langle F \rangle$ is the time-averaged flux of the system.

Measurements of secondary eclipse depth thus constrain a combination of $A_{B}$, $A_{g}$ and $P_{n}$ in this toy model. Observations of the phase function amplitude constrain these same parameters, as well as $\sin i$ and $R_{p}$ in the case of non-transiting systems.

\subsection{Numerical Models}
In addition to the toy model outlined above, there have been numerous more sophisticated attempts to model the atmospheres of Hot Jupiters and thus to predict the depth of secondary eclipses as well as the full phase function of specific extrasolar systems. We summarize the most pertinent aspects of these models here \citep[for a more thorough background see][]{2006ApJ...650.1140B}.   \cite{2005ApJ...629L..45C, 2006ApJ...649.1048C} determine energy recirculation and predict global temperature contrasts of $\sim$500 K at the photosphere. They also predict a noticeable offset between the phase function peak and the time of opposition. 
\cite{2005ApJ...627L..69F} construct models with either no or full recirculation and find that the latter is a better fit to secondary eclipse observations. \cite{2005ApJ...625L.135B}, \cite{2005ApJ...632.1132B} and \cite{2005ApJ...632.1122S} also run simulations with different amounts of recirculation and find that significant redistribution of heat must be occurring to explain the observations. \cite{2006ApJ...652..746F} run models to calculate emitted spectra as a function of orbital phase for both equilibrium and non-equilibrium chemistry. They find that non-equilibrium chemistry leads to significantly smaller phase variations and that current observations can be explained with a blackbody-like planetary spectrum.  Finally, \cite{2006ApJ...650.1140B} use the factor $P_{n}$ to characterize the efficiency of energy recirculation and construct models for five different values of the parameter.  They find that current observations rule out models with no recirculation and are in fact consistent with full recirculation (although they admit that current observations are not precise enough to determine the exact degree of recirculation). 

\subsection{Observational Constraints}
Secondary eclipses have been reported for four short-period extrasolar planetary systems and a phase function has been reported in a different, non-transiting system. We summarize these results below. \cite{2003_Richardson} placed an upper limit of $3\times 10^{-4}$ on the relative secondary eclipse depth at 2.2~$\mu$m using the SpeX instrument at the NASA Infrared Telescope Facility. The secondary eclipse of HD~209458, observed with the Multiband Imaging Photometer for Spitzer (MIPS) yielded a relative depth at 24~$\mu$m of $0.0026\pm0.00044$ \citep{2005Natur.434..740D}. Similarly, the secondary eclipse of TrES-1 was observed in the IRAC 4.5 and 8~$\mu$m bands, yielding eclipse depths of $0.00066 \pm 0.00013$ and $0.00225 \pm 0.00036$, respectively \citep{2005ApJ...626..523C}. The latest system to have been detected in secondary eclipse is HD~189733 \citep{2006ApJ...644..560D}.  Observations with the Spitzer Infrared Spectrograph (IRS) peak-up imager at 16~$\mu$m give an eclipse depth of $0.00551 \pm 0.00030$. In addition to infrared detections, the non-detection of a secondary eclipse for HD~209458 at optical wavebands has constrained the 500 nm secondary eclipse depth to be less than $4.88\times10^{-5}$ at the $1\sigma$ level and less than $1.34\times10^{-4}$ at the $3\sigma$ level \citep{2006ApJ...646.1241R}. Ground-based K-band (2.2~$\mu$m) observations of OGLE-TR-113b tentatively show a secondary eclipse depth of $0.0017 \pm 0.0005$ \citep{2007MNRAS.375..307S}. MIPS 24~$\mu$m imaging of $\upsilon$ Andromeda has yielded a phase function with peak-to-trough amplitude of $0.0029 \pm 0.0007$ and weak evidence of a small phase offset \citep{2006Sci...314..623H}. This surprisingly large amplitude suggests that the planet has parameters $A_{B} \approx 0$, $P_{n} \approx 0$ and $i \ge 30^{\circ}$. Continuous IRAC 8~$\mu$m monitoring of HD~189733 for half of a planetary orbit led to the detection of a phase function with a peak-to-trough amplitude of 0.00121(4) \citep{2007_Knutson}. This modest phase function, roughly one third the relative depth of secondary eclipse, indicates that much of the incident stellar energy is being recirculated to the planet's night side. \cite{2006_Grillmair} used IRS to obtain a mid-IR spectrum for HD~189733b, which showed no evidence for absorption bands due to either water or methane. \cite{2007_Richardson} obtained the relative mid-IR spectrum of HD~209458b and similarly concluded that such absorption features could be excluded; they also identified candidate emission features at two wavelengths. 

\begin{figure}
\includegraphics[width=84mm]{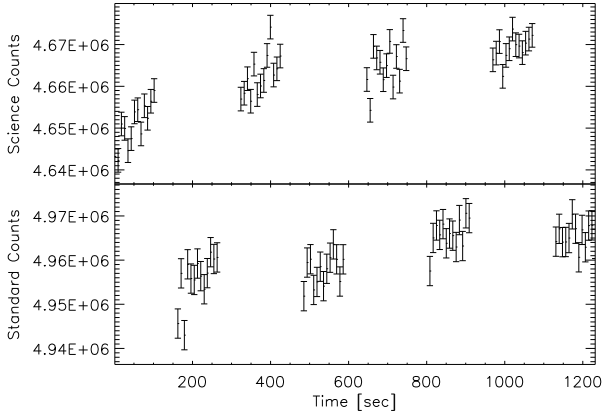}
\caption{Sample 8~$\mu$m observation of science target, 51~Peg, and flux standard, HD~217636, showing the flux ramp-up. This observation began at MJD = 53728.447, or $\phi=1.88\pi$. Each point represents 64 exposures, each of 0.1 sec. The error bars, which have an average value of $3.3\times10^{3}$, correspond to the standard deviation in the counts from the 64 images. For comparison, one would expect  uncertainties of roughly $2.2\times10^{3}$ in the Poisson noise regime.}
\label{51Peg_ch4_observation_2}
\end{figure} 

\section{Description of Observations}
\subsection{Photometry}
We acquired Spitzer photometry\footnote{Cycle 2 Spitzer program GO 20482, PI Agol} with the IRAC 3.6~$\mu$m or 4.5~$\mu$m, and 8~$\mu$m detectors in sub-array mode (32$\times$32 pixels, or 38.3$\times$38.3 arcseconds) to avoid saturation \citep{2004_Fazio}. Observations were made at eight different phases for each planetary system with follow-on time constraints. Each observation consists of four nods to the target star interleaved with four nods to a nearby flux standard of comparable infrared brightness. The number and duration of exposures for each of our targets are summarized in Table~\ref{results}. We convert fluxes from the Spitzer units of specific intensity (MJy/sr) to photon counts, and use aperture photometry to recover the counts for the science and standard stars. We determine the PSF centroid using a centre-of-light function, then use the IDL routine APER to extract the source flux. After testing many combinations of aperture and sky annulus radii for robustness, we settle on a 7 pixel aperture radius and a sky annulus centred on the target star, with an inner radius of 10 pixels and an outer radius of 15 pixels.

\subsection{Calibration}
A single observation lasts only 20 minutes, compared to planetary orbital periods of days, so the flux is not expected to vary noticeably within a single observation.  Unfortunately, the measured signal in the 3.6~$\mu$m and 4.5~$\mu$m bands exhibits large (few \%) fluctuations. The pointing drifts between nods are sub-pixel yet other researchers have found that the IRAC 4.5~$\mu$m InSb array shows significant intra-pixel sensitivity variations \citep{2005ApJ...626..523C, 2006ApJ...653.1454M}. The 3.6~$\mu$m detector, also an InSb array, likely suffers from the same problem. We do not have sufficient images to properly map out the position-sensitivity relation, and this severely limits the quality of our 3.6~$\mu$m and 4.5~$\mu$m data. The 8~$\mu$m observations of 51~Peg and HD~179949 show a smooth increase in gain, with a similar shape for both the science target and flux standard, indicating an instrumental effect, as shown in Figure~\ref{51Peg_ch4_observation_2}. HD~209458, a somewhat fainter source, does not exhibit such a marked ramp-up. The illumination-dependent nature of this instrumental effect, described in greater detail in \cite{2007_Knutson}, is consistent with charge trapping, in that pixel sensitivities are initially deficient but asymptote to a constant value as the pixels are illuminated repeatedly.

For each observation, we determine the total counts by multiplying the median count by the number of frames, thus ignoring the minority of frames affected by cosmic rays. We repeat this median averaging for all the exposures within each nod (with duration $\tau_{nod}$) and take the standard deviation in the counts from each exposure multiplied by the square-root of the number of exposures per nod to be the uncertainty in the total counts. We find these uncertainties to be 4--5 times larger than the expectation from Poisson noise at 3.6~$\mu$m and 4.5~$\mu$m, and roughly twice Poisson noise at 8~$\mu$m. At the shorter wavebands, as well as for the 8~$\mu$m observations of HD~209458, the points from different nods show considerable scatter but no obvious trend, so we add up the counts from all the nods separately for the science and reference star, then take the ratio of the two sums to determine the relative flux at that epoch. Noticing that the ramp-up at 8~$\mu$m for 51~Peg and HD~179949 levels off towards the end of any given observation, we take the ratio of the summed counts for the two final nods as the relative flux at that epoch.  In all cases we compute the uncertainty in the flux ratio by taking the quadratic sum of the relative uncertainties in the science and standard counts. 
The average flux measured for each star can be found in Table~\ref{results}.

\begin{table*}
\begin{minipage}{126mm}
\caption{Number and Duration of Exposures and Total Time on Target, Mid-IR Fluxes and Phase Variability of Target Stars}
\label{results}
\begin{tabular}{@{}lcccccc}
\hline
Star & $\lambda$~[$\mu$m] & $\tau_{\rm exp}$~[sec] & $N_{\rm exp}$ & $\tau_{\rm nod}$~[sec]&$F_{*}(\sigma)$~[mJy] & $\Delta F/ <F>$\\
\hline
\hline
{\bf 51~Peg}  & 4.5  & 0.02 & 1728 & 34.56&533.1(9) &$<0.017$\\
 & 8 & 0.1 & 832 & 83.2& 186.52(8) &$<0.0007$\\
 \hline
HD~217636 & 4.5  & 0.02 & 1728 & 34.56&517.6(9) & N/A\\
 & 8 & 0.1 & 832 & 83.2&  198.2(1) & N/A\\
\hline
{\bf HD~179949} & 3.6  & 0.02 & 2560 & 51.2&331(11) &$<0.019$\\
 & 8 & 0.4 & 576 & 230.4& 73.72(4)&$0.00141(33)$\\
\hline
SAO~187890 & 3.6  & 0.02 & 2560 & 51.2&367(2) &N/A\\
 & 8 & 0.4 & 576 & 230.4& 83.51(5)& N/A\\
\hline
{\bf HD~209458} & 3.6  &0.1 & 320 & 32&92.2(3)&$<0.015$ \\
& 8 & 0.1 & 320 & 32& 20.75(4) &$<0.0015$\\
\hline
HD~210483 & 3.6 &0.1 & 320 & 32&114.1(2) & N/A\\
 & 8 & 0.1 & 320 & 32& 25.49(5) & N/A\\
\hline
\end{tabular}

\medskip
The wavelength of the observations is $\lambda$, $\tau_{exp}$ is the duration of a single exposure, $N_{exp}$ is the number of exposures per nod, $\tau_{nod}$ is the total time on target per nod, $F_{*}(\sigma)$ is the average flux of the system and the associated error, while $\Delta F/F_{*}$ is the $2\sigma$ upper limit, or the best-fit peak-to-trough amplitude and associated $1\sigma$ uncertainty for phase variations, as appropriate.
\end{minipage}
\end{table*}

\begin{figure}
\includegraphics[width=84mm]{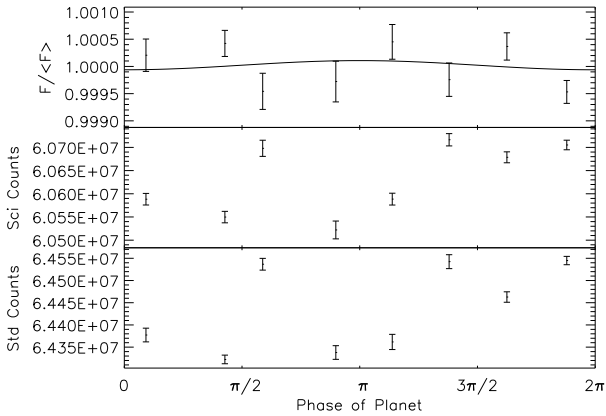}
\caption{The two bottom panels show the 8~$\mu$m light curves for the science target, 51~Peg, and the flux standard, HD~217636. The top panel shows the \textit{relative} light curve (the ratio of the two lower plots). The phase of the planet ($\phi$ from \S1.2) is 0 at inferior conjunction, and $\pi$ at superior conjunction. The $1\sigma$ error bars have an average size of $0.00029$ and represent the standard deviation in the mean of the binned counts. 
The curve is the best-fit sinusoid (with fixed period and no phase offset), as determined by a $10^{5}$-step Markov Chain Monte Carlo.}
\label{51Peg_ch4_light_curve}
\end{figure} 

\begin{figure} 
\includegraphics[width=84mm]{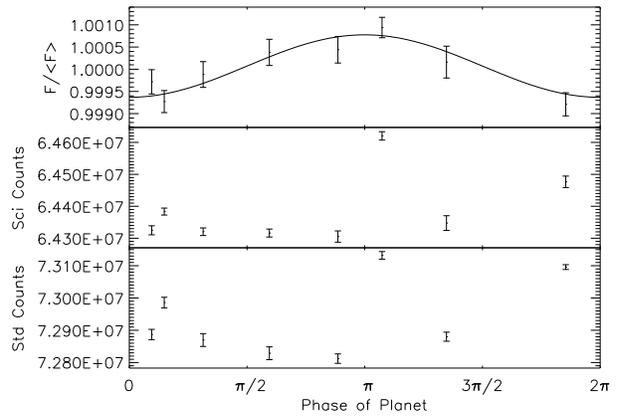}
\caption{The two bottom panels show the 8~$\mu$m light curves for the science target, HD~179949, and the flux standard SAO~187890. The top panel shows the \textit{relative} light curve at for HD~179949.  The phase of the planet ($\phi$ from \S1.2) is 0 at inferior conjunction, and $\pi$ at superior conjunction. The $1\sigma$ error bars have an average size of $0.00028$ and represent the standard deviation in the mean of the binned counts. 
The curve is the best-fit sinusoid (with fixed period and no phase offset), as determined by a $10^{5}$-step Markov Chain Monte Carlo.}
\label{HD179949_ch4_light_curve}
\end{figure} 

\begin{figure}
\includegraphics[width=84mm]{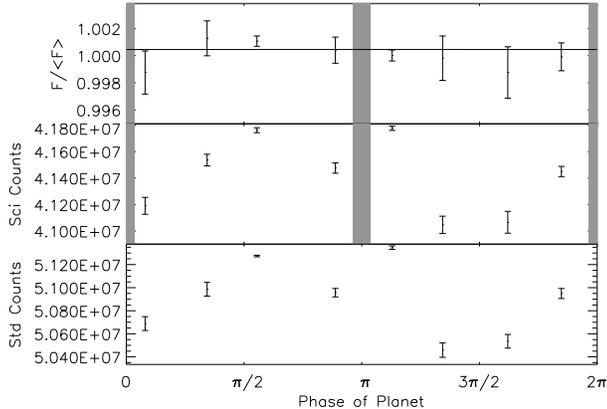}
\caption{The two bottom panels show the 8~$\mu$m light curves for the science target, HD~209458, and the flux standard star, HD~210483. The top panel shows the \textit{relative} light curve.  The phase of the planet ($\phi$ from \S1.2) is 0 at inferior conjunction, and $\pi$ at superior conjunction. The greyed-out areas correspond to a transit or secondary eclipse. The $1\sigma$ error bars have an average size of $0.00070$ and represent the standard deviation in the mean of the binned counts. 
The curve is the best-fit sinusoid (with fixed period and no phase offset), as determined by a $10^{5}$-step Markov Chain Monte Carlo.}
\label{HD209458_ch4_light_curve}
\end{figure} 

\section{Light Curves \& Implications}
We use ephemerides from \cite{2006ApJ...646..505B}, for which projected errors at the epochs of our observations are less than 60 min.\ for all three systems. We neglect the correction for the light travel time across the Solar System, since the size of this correction (less than 8 minutes) is less than the uncertainty in the orbital phase. 
By matching the planet phases with the relative flux at that epoch, we construct a relative light curve for each system, shown in the top panel of Figures~\ref{51Peg_ch4_light_curve}--\ref{HD209458_ch4_light_curve} (the 3.6~$\mu$m and 4.5~$\mu$m light curves are not plotted due to prohibitively large systematic uncertainties). HD~179949 and 51~Peg are non-transiting systems, while the observations of HD~209458 were taken outside of transit or secondary eclipse (represented by shaded areas in Figure~\ref{HD209458_ch4_light_curve}).
 
\begin{figure} 
\includegraphics[width=84mm]{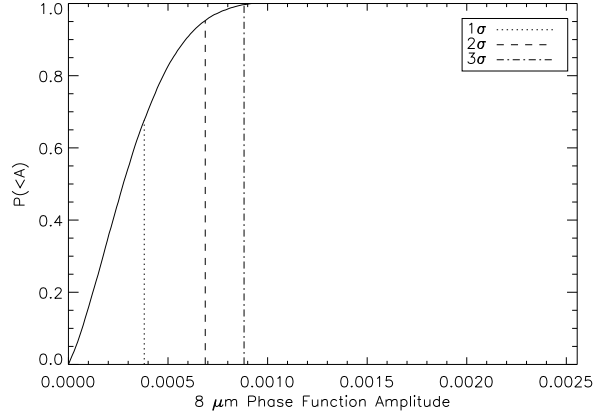}
\caption{Cumulative distribution function for the peak-to-trough amplitude of the phase function of 51~Peg based on Markov Chain Monte Carlo analysis of IRAC 8~$\mu$m images. The x-axis spans the full range of theoretically allowed amplitudes for this system.}
\label{51Peg_ch4_exclusion_plot}
\end{figure}

\begin{figure} 
\includegraphics[width=84mm]{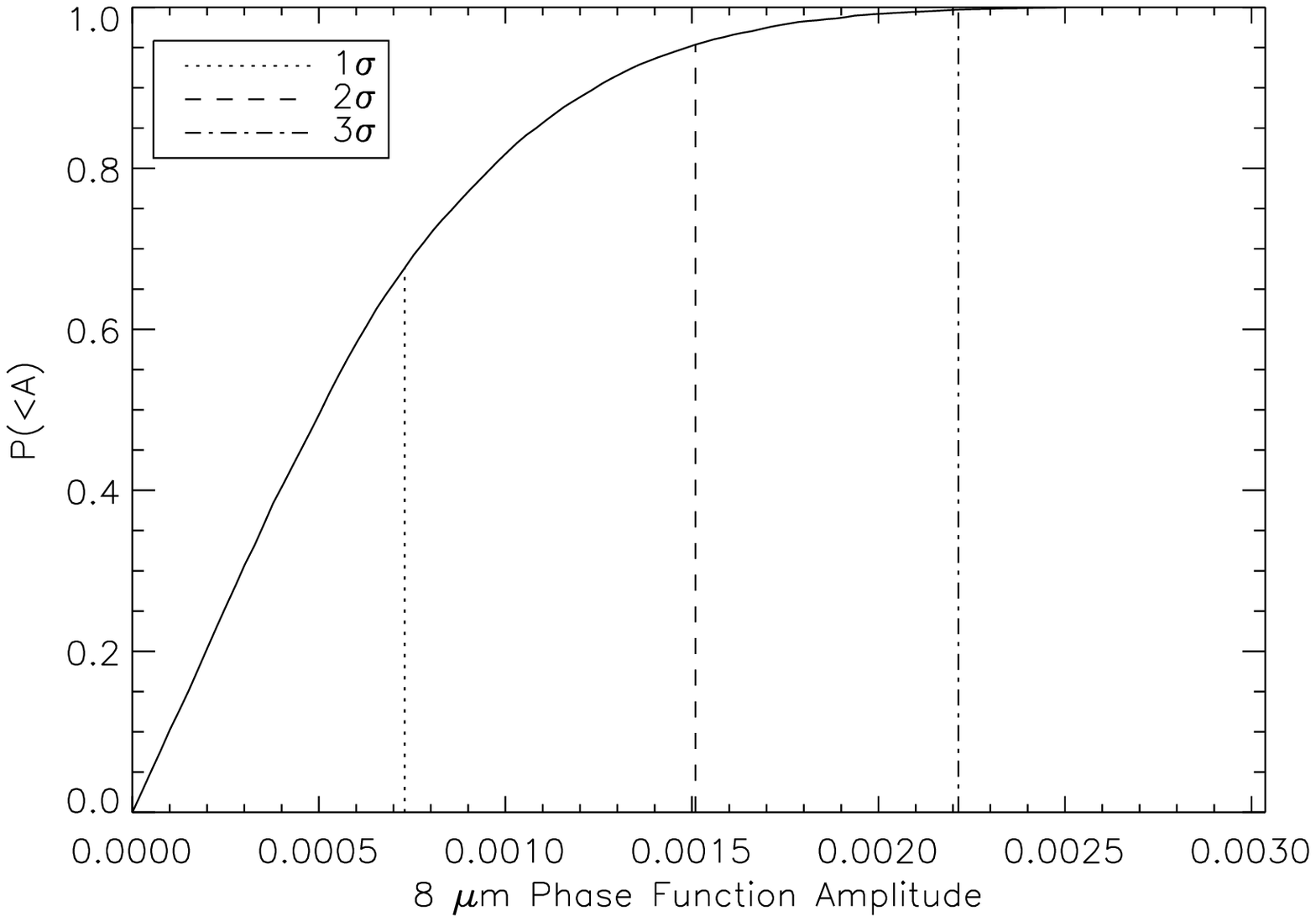}
\caption{Cumulative distribution function for the peak-to-trough amplitude of the phase function of HD~209458 based on Markov Chain Monte Carlo analysis of IRAC 8~$\mu$m images. The x-axis spans the full range of theoretically allowed amplitudes for this system.}
\label{HD209458_ch4_exclusion_plot}
\end{figure}

We use a $10^5$-step Markov Chain Monte Carlo with Metropolis-Hastings algorithm \citep{2005AJ....129.1706F} to determine the best-fit sinusoid, which is shown as the curve in the top panel of Figures \ref{51Peg_ch4_light_curve}--\ref{HD209458_ch4_light_curve}. The amplitude and constant offset of the sinusoid are allowed to vary, while the period is fixed at $2\pi$ and the phase offset is set to zero, in accordance with \S1.2, as well as the results of \cite{2006Sci...314..623H}.
None of the 3.6~$\mu$m and 4.5~$\mu$m light curves, or indeed the 8~$\mu$m light curve for 51~Peg or HD~209458 show a variation that is better fit by a sinusoid than a straight line. We do, however, detect an 8~$\mu$m phase function with peak-to-trough amplitude of $0.00141(33)$ for HD~179949, compared to an amplitude of $0.0031$ we would expect for an edge-on planet with a radius of $1.2R_{J}$ with an albedo of $0$ and a recirculation parameter of $P_{n}=0$.  
For the two systems where we do not detect a phase function, the Markov Chain produces the cumulative distribution functions shown in 
Figures~\ref{51Peg_ch4_exclusion_plot} and \ref{HD209458_ch4_exclusion_plot}. These plots show the full range of theoretically allowed peak-to-trough phase function amplitudes on the x-axis, while the y-axis shows the probability that the system has an amplitude \textit{less than} a given amplitude. The 1, 2 and 3$\sigma$ upper limits are marked for convenience. The $2\sigma$ upper-limits on phase variations at 8~$\mu$m are $0.0007$ for 51~Peg (compared to an amplitude of $0.0026$ for an edge-on planet with $R = 1.2R_{J}$  with an albedo of $0$ and a recirculation parameter of $P_{n}=0$), and $0.0015$ for HD~209458 (compared to an amplitude of $0.0030$ if the planet has an albedo of $0$ and a recirculation parameter of $P_{n}=0$). Our complete results are summarized in Table~\ref{results}. 

We recognize that $\upsilon$ Andromeda and HD~179949 are the only two planet-bearing systems studied by \cite{2005_Shkolnik} to exhibit chromospheric activity in phase with the planetary orbit.  It is therefore possible that the mid-IR phase variations that have been observed in these systems are a consequence of stellar activity and not the diurnal flux contrast of the planet. We do not believe that chromospheric activity is responsible for the phase function we observe in HD~179949, however, as the cycle in chromospheric activity \cite{2005_Shkolnik} observed in this system was $60^{\circ}$ out of phase with the planetary orbit, and this pattern persisted for at least the year they observed it.
 
\begin{figure} 
\includegraphics[width=84mm]{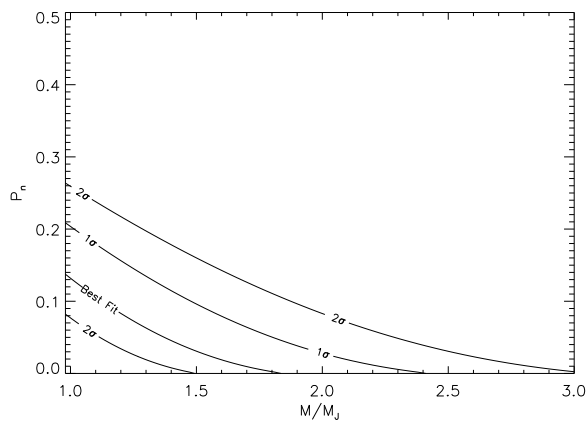}
\caption{Exclusion plot for the mass, $M/M_{J}$, and recirculation, $P_{n}$, of  HD~179949b based on its 8~$\mu$m phase function. $P_{n}=0$ corresponds to no redistribution, while $P_{n}=0.5$ corresponds to full redistribution. Lower masses correspond to inclinations near $\pi/2$, while greater masses correspond to more face-on configurations. The radius of planet is assumed to be in the range $R_{J} < R_{p} < 1.2R_{J}$, and its Bond Albedo is set to $0$. The $1\sigma$ and $2\sigma$ confidence intervals for large phase function amplitudes lie on top of each other so the former is omitted for clarity.}
\label{HD179949_ch4_exclusion_plot_2}
\end{figure}

If the phase variation of HD~179949 is taken at face value, it constrains the properties of this system, as explained in Section~\ref{toy}.  Constraining the planetary radius to within $R_{J} < R_{p} < 1.2R_{J}$, the observed phase variation places a $1\sigma$ upper-limit of $P_{n} < 0.21$. Further constraining the albedo to $A_{B}=0$, consistent with the value of $A_{g}=0.03(5)$ that \cite{2007_Rowe} measured for HD~209458b, we obtain the exclusion plot shown in Figure~\ref{HD179949_ch4_exclusion_plot_2}. Our model thus places a $1\sigma$ limit of $M<2.4 M_{J}$ on the mass of HD~179949b.

\begin{figure}
\includegraphics[width=84mm]{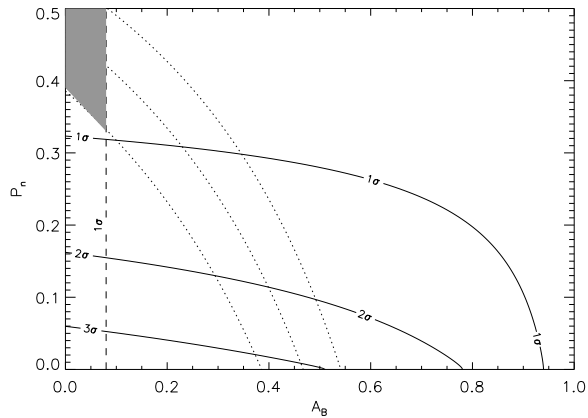}
\caption{Exclusion plot for the Bond albedo, $A_{B}$, and recirculation, $P_{n}$ of  HD~209458b based on its 8~$\mu$m phase function. $P_{n}=0$ corresponds to no redistribution, while $P_{n}=0.5$ corresponds to full redistribution. Solid lines correspond to upper limits in this work. The dotted lines corresponds to constraints from the secondary eclipse measurement (D. Charbonneau, personal communication), while the dashed line is from the \protect\cite{2007_Rowe} albedo upper limit. The bottom-left corner of the plot (the no albedo, no recirculation configuration) is excluded at the $3\sigma$ level by our data, while the shaded region in the top left of the figure is within all the $1\sigma$ contours.} 
\label{HD209458_ch4_exclusion_plot_2}
\end{figure}

Because of the edge-on configuration of HD~209458, our upper-limit can be combined with the constraint on the planet/star radius ratio of \cite{2002_Mandel}, yielding constraints on the albedo and recirculation of the Hot Jupiter's atmosphere (s.f. \S~\ref{toy}). In particular, the $A_{B} = 0$, $P_{n}=0$ configuration of our toy model is excluded at the $3\sigma$ level, as shown in Figure~\ref{HD209458_ch4_exclusion_plot_2}. More importantly, we exclude the much more realistic atmospheric models of \cite{2006ApJ...650.1140B} for $P_{n} \le 0.4$ by more than $2\sigma$, since their models predict stronger phase variations than our toy model for the same $P_{n}$. 
 
\section{Conclusions}
We have acquired Spitzer IRAC 3.6~$\mu$m or 4.5~$\mu$m, and 8~$\mu$m images of three known Hot Jupiter systems. The use of standard stars has allowed us to achieve better than 0.05\% relative photometry with multi-epoch IRAC 8~$\mu$m observations. Intra-pixel sensitivity variations limit our photometry at 3.6~$\mu$m and 4.5~$\mu$m.  We detect a 8~$\mu$m phase function for HD 179949 and determine upper limits for the amplitude of phase variations in the two other observed systems. We analyze these light curves using a toy model of the planets and constrain the properties of the planets.  Our observations suggest that Hot Jupiters do not all exhibit the large mid-infrared diurnal flux contrast detected by \cite{2006Sci...314..623H}. Even within our modest sample of three planets, it appears that HD~179949b and HD~209458b exhibit rather different degrees of recirculation. 
The data for HD~179949 and $\upsilon$ Andromeda provide the most uncertain constraints, however, while the 
two transiting systems with well-constrained phase functions \citep[HD~209458 and HD~189733, ][]{2007_Knutson} both point 
toward a large value of $P_{n}>0.35$, so it remains to be seen if the observations of the non-transiting planets hold up under further scrutiny.

\section*{Acknowledgments}
The authors would like to acknowledge Mark Claire and Daryl Haggard for their insightful comments, and Greg Laughlin for his help tracking down ephemerides. N.B.C.\ has been supported in this research by the Natural Sciences and Engineering Research Council of Canada, le Fonds qu\'eb\'ecois de la recherche sur la nature et les technologies, and the U.W.\ Astrobiology Program. This work is based on observations made with the Spitzer Space Telescope, which is operated by the Jet Propulsion Laboratory, California Institute of Technology under a contract with NASA. Support for this work was provided by NASA through an award issued by JPL/Caltech.

\end{document}